\documentclass[a4paper]{jpconf}
\usepackage{graphicx}

\begin{document}

\title{Microturbulence studies in RFX-mod}

\author{F Sattin${}^1$, S C Guo${}^1$, I Predebon${}^1$, S F Liu${}^2$, X Garbet${}^3$ and M Veranda${}^1$} 
\address{${}^1$ Consorzio RFX, Associazione Euratom-ENEA per la fusione, Padova, Italy}
\address{${}^2$ Department of Physics, Nankai University, Tianjin, People’s Republic of China}
\address{${}^3$ CEA, IRFM, F-13108 Saint Paul Lez Durance, France}

\ead{ fabio.sattin@igi.cnr.it}

\begin{abstract} Present-days Reversed Field Pinches (RFPs) are characterized by quasi-laminar magnetic configurations in their core, whose boundaries feature sharp internal transport barriers, in analogy with tokamaks and stellarators. The abatement of  magnetic chaos leads to the reduction of associated particle and heat transport along wandering field lines. At the same time, the growth of steep temperature gradients may trigger drift microinstabilities. In this work we summarize the work recently done in the RFP RFX-mod in order to assess the existence and the impact upon transport of such electrostatic and electromagnetic microinstabilities as Ion Temperature Gradient (ITG), Trapped Electron Modes (TEM) and microtearing modes.    
\end{abstract}

\section{ Introduction}
The current paradigm for high-performance discharges in Reversed Field Pinches (RFPs) is provided by SHAx states: magnetic equilibria characterized by field lines twisting around one helically wound axis. These states feature well conserved magnetic surfaces in the core; the possibility of their existence was theoretically proposed several years ago \cite{uno}, experimentally identified more recently \cite{due,tre,quattro}, and represent nowadays the regular working scenario, at least in the RFX-mod device, to which we will refer throughout this paper.

The border of the magnetically ordered volume is characterized thermally by steep internal energy transport barriers (ITBs). The region of high magnetic chaos and consequently degraded transport is pushed to the edge, downhill to the ITB\footnote{Note that--just like tokamaks--RFPs feature edge transport barriers as well \cite{cinque}. These barriers are not subject of the present study}. Remarkable progresses have been done with respect to the earlier chaotic RFPs as far as energy losses are concerned, but power-balance estimates of the heat conductivity suggest that its value is still above the collisional level \cite{quattro}: either some residual overlapping of MHD tearing modes provides non-negligible transport of magnetic nature, or other instabilities are at work. An extensive coverage of the actual status of knowledge about the physics of present-days RFPs is provided by Cappello{\it et al} \cite{sei}. Recent theoretical research has focused on investigating both possibilities: as far as the topological characterization of the core from the MHD point of view is concerned, see the work by Bonfiglio \textit{et al} presented at this conference \cite{sette}. In this work, we will investigate the possible existence and role of microinstabilities in SHAx states, how and whether these instabilities could be triggered, and to what extent they could affect overall transport. We aim at providing a brief survey of the results of this activity obtained at RFX-mod along the past two years. We supplement existing results with novel unpublished material; furthermore, hints to new lines of research as well as to issues not yet completely settled, will be given. The outline of the work is as follows: In section 2 Ion Temperature Gradient (ITG) modes will be addressed. Their stability properties will be examined in connection with pure electron-ion (“pure”) plasmas (Sec. 2.1), and then possible modifications due to plasma pollution by impurities will be considered for (Sec. 2.2). Section 3 is devoted to studies of Trapped Electron Modes. Finally, an extensive coverage is provided in Sec. 4 of Microtearing Modes. On the basis of current evidence, these modes appear likely candidates as possible sources of residual turbulent transport at the location of ITBs. 

\section{Ion Temperature Gradient Modes}
\subsection{ ITGs in pure plasmas} 
ITGs are regarded as one of the main sources of turbulent transport in tokamaks; accordingly, they were the first to be addressed by investigations. It was early noticed by Guo \cite{otto} (see also \cite{nove}) that RFPs are more resilient than Tokamaks to this kind of instability. The rationale for this behaviour stays in the different level of Landau damping due to the different connection length $L_c$ of magnetic field lines between the two devices. Waves with helicity defined by the angular numbers $(m,n)$ are damped through energy exchange to those particles that are close to the resonance condition
\begin{equation}
 v_{||} = \omega k_{||}^{-1} \quad k_{||} = 2 \pi \times (m - n q)\times  L_c^{-1}
\end{equation}		
where $v_{||}$ is the parallel (to the magnetic field) component of the particle velocity, $k_{||}$ the parallel wavenumber of the mode, $\omega$ its angular frequency, and $q$ the safety factor. In RFPs, by virtue of their geometry, $L_c$ is shorter than in Tokamaks by about a factor (minor radius)/(major radius), accordingly, $k_{||}$ is larger by the same amount. Landau damping of the wave is the most effective the more resonant particles are present. If the velocity distribution function $f(v_{||})$ is close to a Maxwellian, the smaller $v_{||}$ the more resonant particles are available. As a consequence, the critical temperature scale length   needed for triggering the instability turns out to be substantially shorter than in Tokamaks. Earlier semianalytical calculations where later reinforced by full numerical linear electrostatic gyrokinetic calculations using GS2 code \cite{dieci}, suitably adapted to account for RFP magnetic topology \cite{undici}: under common plasma conditions, it may be stated that ITG instability at mid-radius arises for $R/L_{Ti} > 20 \div 25$   ($R$ being the major radius): at least a fourfold factor larger than for similar plasma conditions in tokamak geometry (Fig. 1). An independent investigation using linearly the code TRB (adapted to RFP geometry) led to the same quantitative result \cite{dodici}\footnote{Despite being a fluid code, TRB may account for such a kinetic effect as the Landau damping by implementing the Hammett-Perkins (H-P) closure scheme \cite{tredici}. Thus, agreement between TRB’s and GS2 fully kinetic treatment, provided as a side effect further confidence about the reliability of the H-P approximation.}: see Fig. (1), where TRB result (the large circle) lies quite accurately on top of GS2 ones.

\begin{figure}
\includegraphics{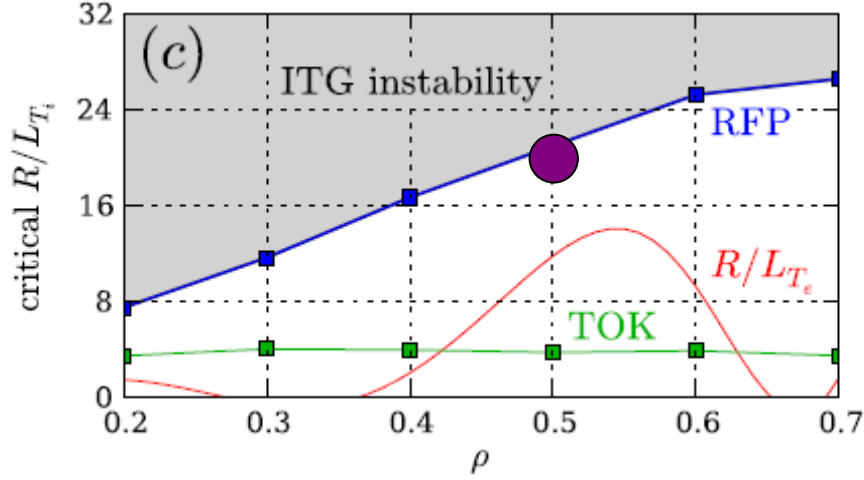}
\caption{ITG stability threshold versus normalized minor radius $\rho=r/a$ for typical RFX-mod plasma conditions (RFP) and for a tokamak profile (TOK). The threshold is the $R/L_{Ti}$ value at which the growth rate become positive. The red curve is the typical experimental value of $R/L_{Te}$. The blue curve with symbols are GS2 results for the RFP. The violet dot is the equivalent quantity computed by TRB. The green curve with symbols is the stability threshold for a tokamak with the same plasma parameters. }
\end{figure}

In Fig. (2) an instance of a nonlinear time-dependent TRB simulation is presented, out of a yet preliminary set of simulations aiming to investigate a possible sinergy between the ITG-driven and other sources of  turbulence in SHAx discharges. An initially ITG-unstable profile (red curve) is let to evolve on top of a fixed background diffusivity, $\chi_{back}$ , that attempts to modelize the residual MHD tearing-mode driven diffusivity, and is chosen small in the inner half radius and large in the outer half. The total heat conductivity is thus $\chi = \chi_{back} + \chi_{ITG}$, where the former term is known and fixed, and the latter is self-consistently calculated by TRB. In principle, the existence of a irreducible level of conductivity might enforce temperature gradients, and thus provide a constant free energy source for triggering ITGs. However, within the numerical range of diffusivites scanned--still rather preliminary--we did not find evidence for such mechanism: all curves collapsed towards practically the same profile, basically driven just by ITGs.      

\begin{figure}
\includegraphics[width=8cm]{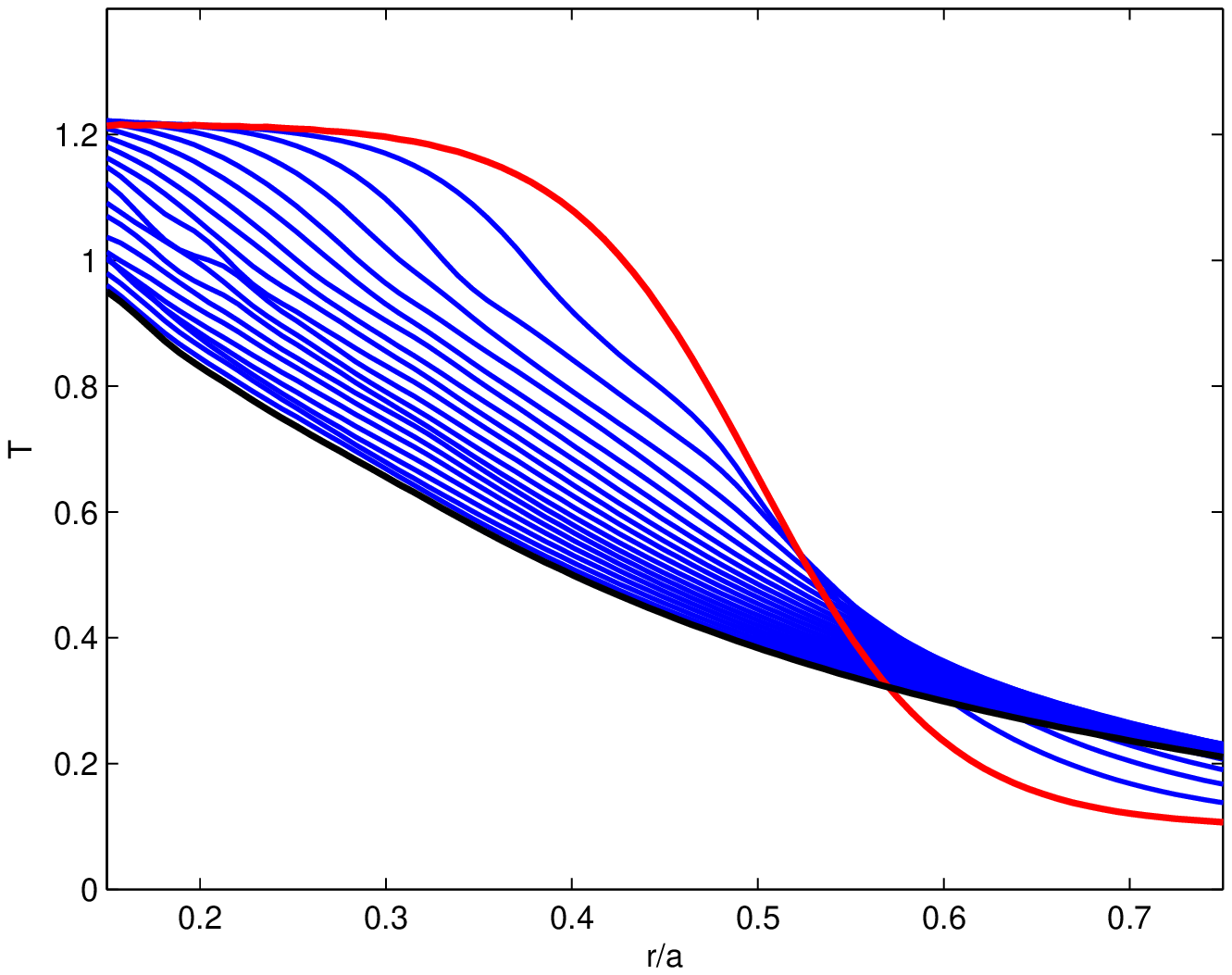}
\includegraphics[width=8cm]{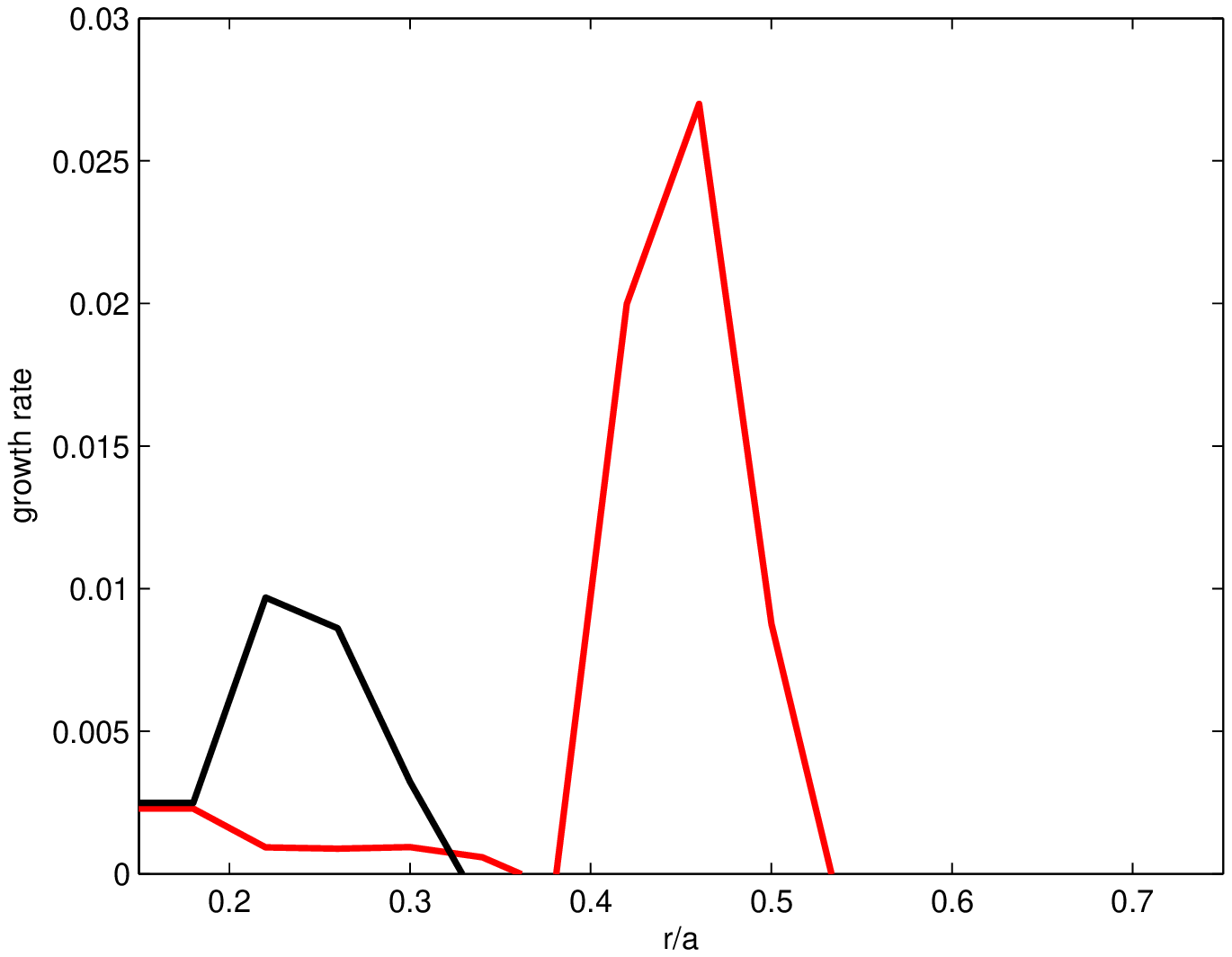}
\caption{Left panel, an initially ITG-unstable ion temperature profile (red curve) is let to evolve (blue curves) under self-generated turbulent transport. The black curve is the final profile computed in the numerical simulation. It is quite close to the asymptotical stable profile, as it can be guessed from the right panel: a linear study of both the initial and final temperature profiles shows that in the former case the maximum growth rate is positive, whereas the final profile is close to marginal stability (red and black curves respectively). 
In these simulations the ratio $T_i/T_e$ is held fixed. The heat source is kept constant throughout the whole radius: it is a simplifying but not really satisfactory assumption in RFPs, where the heating is purely ohmic. Finally, the $q$ profile is held fixed, too.  }

\end{figure}

An accurate comparison between codes’ predictions and experimental profiles is not easy to be done, in part because some of the information needed by codes is still not available with accuracy from measurements in RFX-mod: the safety factor $q$, which is reconstructed by an equilibrium code based upon measurements by external probes, and the ion temperature profile, which is inferred from spectroscopic measurements of impurities \cite{quattro}, but not actually measured. Furthermore, if ITG modes are due to be the main drive of transport, one expects to encounter only marginally stable situations, thus growth rates close to zero. With allowance for the uncertainty in some plasma conditions, the present evidence in RFX-mod favours the view that ITGs are probably sub-critical in most discharges, although it cannot be ruled out the possibility that—at the location of particularly steep ITBs—marginal instability conditions may be attained (see, e.g., \cite{otto}). Accordingly, we may argue that ITGs might become of future concern for RFPs, should the present trend towards steeper ITBs continue. 

\subsection{ ITGs and impurities: preliminary results}
RFX-mod plasmas are ordinarily polluted by intrinsic carbon and oxygen impurities, yielding $Z_{eff} \ge 2$ throughout most of the radius. This raises the question whether impurities may play a role in (de)stabilizing ITGs.  The conclusion from the analysis of the linear stability properties of the mode by the gyrokinetic integral eigenmode equation, is that impurities are destabilizing for ITG modes\footnote{This is partially at odds with the Tokamak case, where impurities are generally stabilizing for ITGs, and destabilizing for TEMs (see, e.g., \cite{diciasette}).}  at the locations where $ L_{eZ} = L_n/L_Z < 0$; that is, the impurity profile must be hollow where plasma density profile is concave, and viceversa. The rationale for this result relies yet on Eq. (1): the mode frequency is increased as  $- L_{eZ}$ increases towards larger positive values. For a given wavenumber $k_{||}$  this change acts to shift the resonance condition (1) to a position in velocity space lesser populated, and thus diminish the impact of the Landau damping.  
This study is obviosly highly simplified, being carried out for just one impurity. Actually, each impurity in its several charge states should be accounted for. It is known that, in RFX-mod, impurities do not accumulate in the core, and peak quite outwards \cite{quattordici}. Therefore, it is quite likely that a region where the criterion $L_{eZ}  < 0$ be fulfilled does exist. 

\begin{figure}
\includegraphics[width=7.1cm]{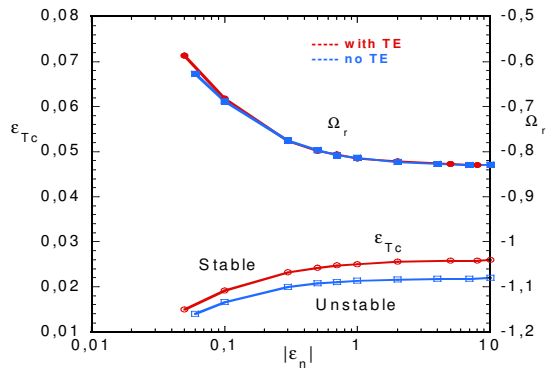}
\caption{Comparison of the ITG threshold $\varepsilon_{Tc} = L_{T(critical)}/R$ and corresponding real frequency $\Omega_r$ (taken negative) between the cases with and without TE. The x axis is  $\varepsilon_n=L_n/R$.  Curves have been computed at fixed wavenumber: $k_y \rho_i \approx 0.45.$}
\end{figure} 

\section{Trapped Electron Modes}   
The fraction of trapped electrons (TE) in a RFP is almost equal to its Tokamak counterpart throughout most of the volume, and somewhat smaller only in the outermost region \cite{quindici}, thus TE could potentially provide as large a contribution as in tokamaks in destabilizing turbulence. Two possible effects of TE were investigated. (I) First of all, there is the possible coupling with ITGs, with destabilizing effects, since TE-driven modes resonate at the same wavelengths as ITGs. However, studies both with TRB (by switching off the TE fraction) and GS2 \cite{sedici} showed that threshold of the mode is not quantitatively altered by their presence (Fig. 3). (II) Further analytical results show that, instead, Trapped Electron Mode (TEM) instability can arise in a RFP, essentially driven by strong density gradients (Fig. 4): actually, under the limit $R/L_n >> 1 $ (where $L_n$  is the density scale length), GS2 simulations show that TEMs are the dominant electrostatic instability. However, it is important to recall that present RFPs do not feature ordinarily core peaked density profiles, since there is not appreciable central particle source. Therefore, the present results do not apply to ordinary plasma conditions, rather to non-standard conditions attained, e.g., transiently by means of pellet injections.      

\begin{figure}
\includegraphics[width=7cm]{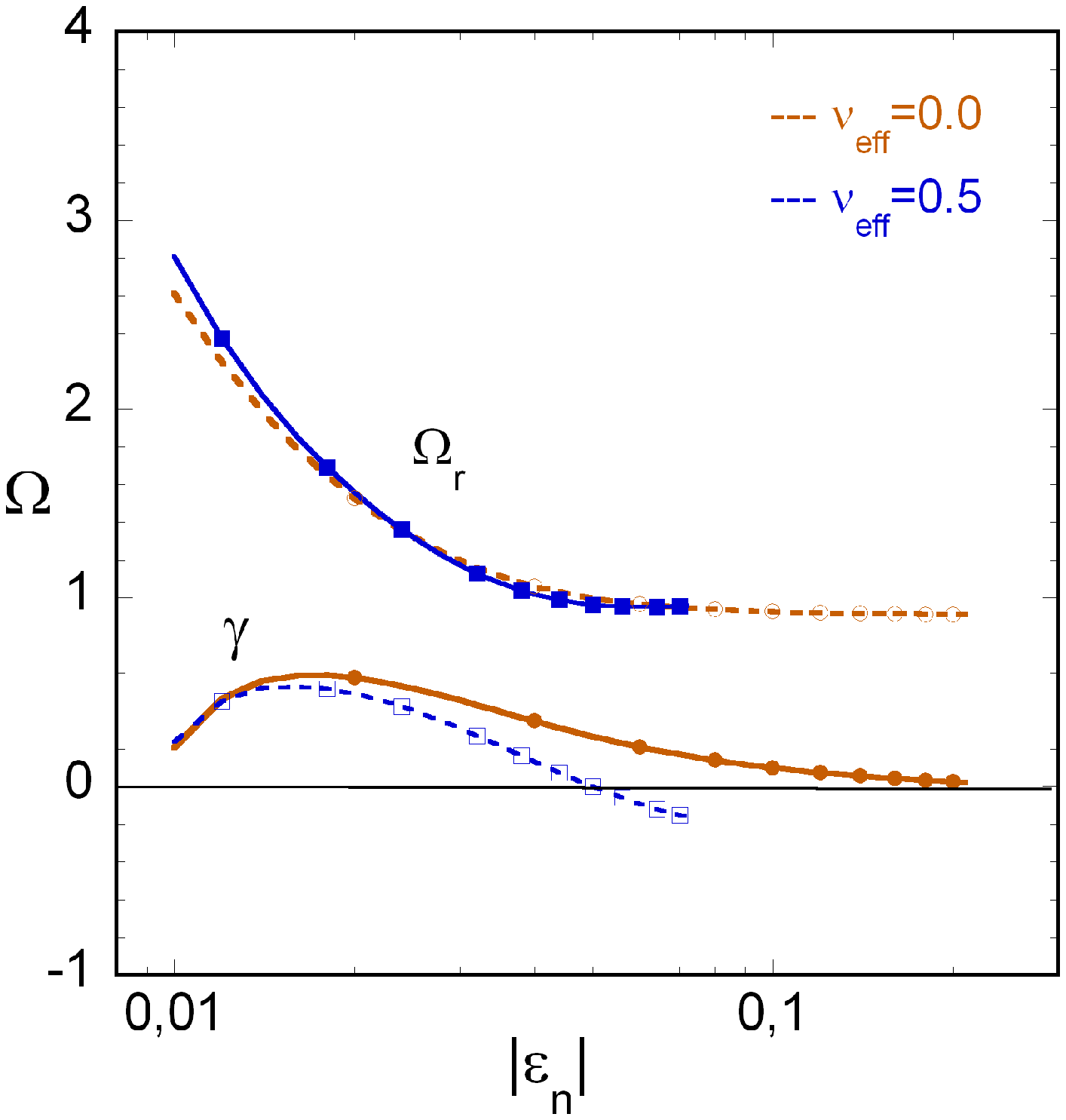}
\caption{The frequency $\Omega_r$ and the growth rate $\gamma$ of TEM are plotted as functions of $|\varepsilon_n |$ for different normalized collision frequencies ($\nu_{eff} = \nu R q/c_s$). In this plot, $R/L_{Te} = 0.025$ and $R/L_{Ti} \approx 0.$}
\end{figure}

\section{Microtearing Modes}
\subsection{ Linear growth rates}
The Microtearing Mode (MT) is a high-wavenumber drift-tearing mode driven linearly unstable by electron temperature gradients, unlike its long-wavelength counterpart, which is essentially current-driven. The growth of MT may lead to chains of overlapping magnetic islands and subsequent local stochastization of magnetic field lines near mode rational surfaces. Hence, it was speculated in the ‘70s that MTs might provide an effective contribution to the thermal diffusivity through electron parallel motion along stochastic field lines. However, since it was also argued that, when electron collision frequency decreases substantially below the electron diamagnetic frequency these modes become stabilized \cite{diciotto}, and most present-days tokamaks  are effectively collisionless in the core, the study of MTs was relegated to the edge region. More recently, interest in these modes was revived in connection with ITBs, where large temperature gradients are sustained (see, e.g., \cite{dicianove}). RFX-mod features presently a moderate peak temperature ($1 \div 1.5$  keV) which, coupled to the strong internal temperature gradients, should represent an optimal environment for these modes to grow. 
Simulations were carried out using GS2: linear runs, with fluctuations of the magnetic vector potential in both parallel and perpendicular components included. The analysis presented is based on the profiles of one reference experimental SHAx case (shot 23977). For such discharge the collision frequency and the electron diamagnetic frequency stay roughly in the range $\nu \approx 10 \times \omega^*$. The dominant core instability turns out to be of MT type (Fig. 5), essentially driven by the electron temperature gradient. The tearing nature is revealed by the parity of the mode, odd for the electrostatic potential and even for the parallel magnetic vector potential. In Fig. (6) mode frequency and growth rate are plotted versus the wavenumber, for some radial positions.

\begin{figure}
\includegraphics[height=6cm]{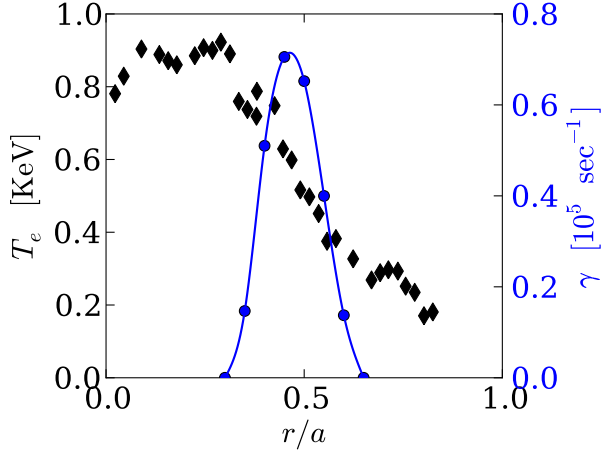}
\caption{Electron temperature profile (diamonds), and the growth rate of the most unstable MT mode (solid curve with circles). }
\end{figure}

\begin{figure}
\includegraphics[height=6cm]{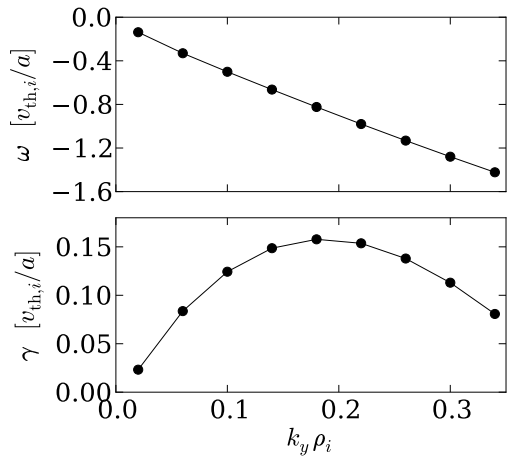}
\caption{Mode real frequency (upper plot) and growth rate (lower plot) of the most unstable mode as a function of the wavenumber, at mid-radius. Frequencies are normalized to (ion thermal speed)/(minor radius). Note that here the sign convention opposite to that of figs. 3,4 has been used: the ion diamagnetic drift has the positive sign.}
\end{figure}

MTs thus appear to provide a sensible contribution to turbulent transport, at the location of the ITB, where$ R/L_T$ is maximum. The question naturally arises: quantify the MT-related transport in terms of an effective heat conductivity in present plasma conditions.

\subsection{ Heat diffusivity due to MTs}
A self-consistent calculation of the MT-driven heat flux using GS2 turned out unfeasible due to the heavy computational requirements for the nonlinear run. We adopted therefore the quasi-linear estimate 
 \begin{equation}
\chi \approx \tilde{b}^2 u_{th,e} L_{corr}
\end{equation}
Inside this expression, the only quantity known accurately is the electron thermal speed $u_{th,e}$. The normalized amplitude of magnetic field perturbations is estimated according to the criterion suggested by Drake {\it  et al}  \cite{diciotto}:  $\tilde{b} \approx \rho_e /L_T, \rho_e$ being the electron Larmor radius. The longitudinal correlation length $L_{corr}$ is usually taken in tokamaks equal to the connection length, however--to the best of our knowledge--no checks have been done to ensure that this expression makes sense in RFPs as well. Thus, we resorted to direct numerical evaluation of $ L_{corr}$: a sinthetic magnetic field was built by superposing a large population of modes with different wave-numbers $(m, n)$ and with amplitude such that their cumulated effect yields a field amplitude of order $\tilde{b}$ . A plot showing the pattern of modes taken into account is given in fig. (7a). A large number of field lines was tracked using the field-line-tracing code NEMATO \cite{venti}. The resulting field topology turns out to be fairly stochastic (Fig. 7b).  Finally, $ L_{corr}$ is estimated as
\begin{equation}
L_{corr} = \int_0^{\infty} { <b(l) b(0)> \over <b^2(0)>} dl
\end{equation}
In this expression, $b$ is the amplitude of the magnetic field, $ l$ the distance travelled along the trajectory of the field line, and the average $<>$  is taken over the ensemble of field lines. We recovered that  $L_{corr} \approx 2 \pi a$. This figure is expected on the basis of the n\"aive identification (correlation length = connection length); on the other hand, at first sight it is a bit puzzling that it does not depend upon the typical wavelength of the magnetic field perturbations involved, as one can argue by a comparison with the related work by D’Angelo and Paccagnella \cite{ventuno}.

\begin{figure}
\includegraphics[height=6cm]{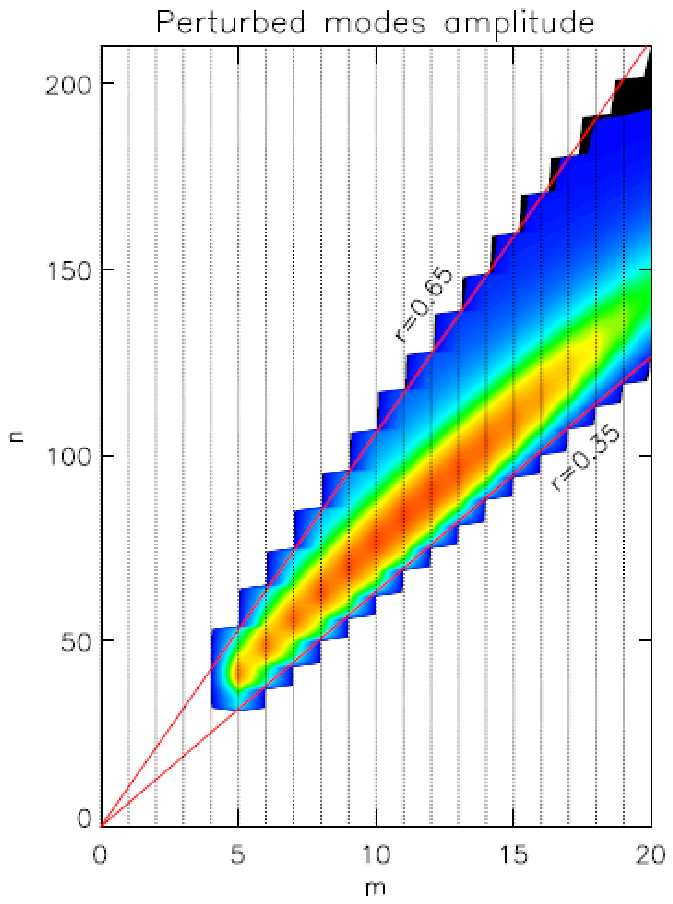}
\includegraphics{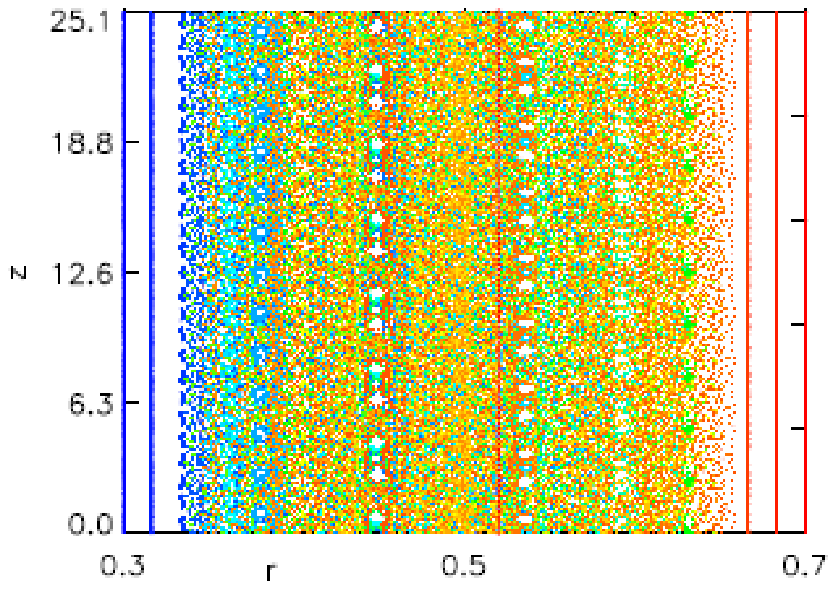}
\caption{Left plot, Perturbed modes’ amplitude contour plot. Right plot, Poincar\'e plot (in the radial-toroidal plane) built from tracking a large number of field lines. Different colors label different starting positions. }
\end{figure}

We could eventually estimate Eq. 2; fairly interestingly, it turned out  $\chi \approx 5\div 20 $ m${}^2$/s: the precise figure depends upon plasma properties as well as upon the detailed way as the perturbed field is assembled\footnote{It is interesting to notice that D’Angelo and Paccagnella \cite{venti} performed an exercise similar to the present one by using a magnetic field perturbation rather different from the present one, and recovered similar figures for the conductivity.}. This figure is fairly compatible with independent power-balance estimates of $\chi$ at the ITB, that yield   $5\div 50  $ m${}^2$/s \cite{quattro}. This strongly supports the view that a relevant fraction of the turbulent transport across the ITB is driven by MTs. Note however that it is not the only transport source: this is suggested by the fact that we found definitely positive growth rates in correspondence of real plasma profiles. If transport was mainly determined by MTs, profiles would accommodate along marginal stability curves. In particular, pressure-driven resistive g-modes have been experimentally identified in the outer plasma region, and might contribute even more internally \cite{ventidue}.

\section{Conclusions}
The RFP, even in the new quasi-laminar SHAx state, confirms to be a device essentially prone to electromagnetic tearing instabilities, although now localized on the microscopic ion Larmor radius scale. Conversely, it appears to be resilient to purely electrostatic modes (ITGs and TEMs): these modes are found potentially unstable, but they are so in plasma conditions characterized by strong gradients (either temperature or density gradients) which are not common in present operating conditions, although intrinsic impurities may affect the extent of this assertion.     

\ack
S. Cappello, D. Escande, E. Fable and R. Paccagnella helped improving the manuscript. This work was supported by the European Communities under the contract of Association between EURATOM/ENEA. The views and opinions expressed herein do not necessarily reflect those of the European Commission.

\end{document}